 \documentclass[11pt]{article}
 \usepackage{fullpage}

\usepackage{amssymb,amsmath,amsfonts}
\usepackage{amsmath}
\usepackage{graphicx}
\usepackage{times}
 \usepackage{ifpdf}
 \ifpdf\fi
\usepackage{float}

\usepackage{algorithm}
\usepackage{algorithmicx}
\usepackage[noend]{algpseudocode}

\usepackage{epsfig}
 \usepackage{times}

\newcommand{\ignore}[1]{}
\newcommand{\signore}[1]{}

\newcommand{\notinproc}[1]{}

\newtheorem{thm}{Theorem}[section]
\newtheorem{theorem}{Theorem}[section]

\newtheorem{lemma}[thm]{Lemma}

\newfloat{fexample}{thp}{lop}
\floatname{fexample}{Example}

 \newcommand{\qed}{\hfill \rule{1ex}{1ex}\medskip\\}
 \newenvironment{proof}{\paragraph{Proof}}{\qed}

\def\range{\mbox{{\sc rg}}}

\def\E{{\textsf E}}

\def\vecv{\boldsymbol{v}}
\def\vecz{\boldsymbol{z}}

\def\L{L*}

\def\1DE{1DE}

\begin{document}

\title{Variance Competitiveness for Monotone Estimation:\\ Tightening
  the Bounds}

\date{}

 \ignore{
\numberofauthors{1}
\author{
\alignauthor Edith Cohen\\
       \affaddr{Microsoft Research SVC}\\
       \affaddr{Mountain View, CA, USA}\\
       \email{editco@microsoft.com}
}
 }

\author{
Edith Cohen\thanks{Microsoft Research, Mountain View, CA, USA
{\tt editco@microsoft.com}} 
}



 \maketitle
\begin{abstract}
 { \small
  Random samples are extensively used to summarize massive data sets
and facilitate scalable analytics.  Coordinated sampling, where
samples of different data sets ``share'' the randomization, is a
powerful method which facilitates more accurate
estimation of many aggregates and similarity measures.

  We recently formulated a model of {\em Monotone Estimation Problems} (MEP),
which can be applied to coordinated sampling, projected on a single item.  MEP
estimators can then be used to estimate sum aggregates, such as
distances, over coordinated samples.
For MEP, we are interested in estimators that are unbiased and nonnegative.
  We proposed  {\em variance competitiveness} as a quality
  measure of estimators: For each data vector, we consider the
 minimum variance attainable on it by an unbiased and nonnegative estimator.
We then define the competitiveness of an estimator as the maximum
ratio, over data,  of the expectation of the square to the minimum
possible.
We also presented a general construction of the
L$^*$ estimator, which is defined  for any MEP
for which a nonnegative unbiased estimator exists, and is at most 4-competitive.

  Our aim here is to obtain tighter bounds on the
{\em universal ratio}, which we define to be the smallest
competitive ratio that can be obtained for any MEP.
  We obtain an upper bound of 3.375, improving over the bound
of $4$ of the \L\ estimator.  We also establish a lower bound of 1.44.
The lower bound is obtained by constructing the {\em optimally
  competitive} estimator for particular MEPs.
 The construction is of independent
interest, as it facilitates estimation with instance-optimal competitiveness.
 }
 \end{abstract}


\section{Introduction}

   We consider sampling schemes where the randomization is captured
   by a single parameter, $u$, which we refer to as the {\em seed}.
   The sampling scheme is specified by a data domain ${\bf V}$,
the {\em seed} $u\in [0,1]$, and a function $S(u,\vecv)$, which maps a data
   $\vecv$ and a value of $u$ to the {\em sample} (or the outcome).
 For each outcome $S$ (and seed $u$), we can consider the set
 $$S^*=\{\vecz\in{\bf V} \mid S(u,\vecz)=S(u,\vecv) \}$$ 
of all data vectors $\vecv\in {\bf V}$, which are consistent with $S$.
 The set $S^*(u,\vecv)$ must clearly include $\vecv$, but can include many, a possibly infinite
   number of data vectors.  

   We recently introduced a  framework of monotone sampling and
   estimation \cite{sorder:PODC2014}, which we briefly review and
   motivate here.

  We say that the sampling scheme of the form above is {\em monotone}  if
   for all $\vecv$, the function $S^*(u,\vecv)$ is non-decreasing with
   $u$.   The sample $S$ can be interpreted as a lossy measurement of the data $\vecv$, which provides some
  partial information, where the seed $u$ is the granularity
  of our measurement instrument, the lower $u$ is, the more we know on
  the data.

  A {\em monotone estimation problem} (MEP)  is defined by a monotone sampling
  scheme together with a function $f:{\bf V} \geq 0$, which we would
  like to estimate.
The estimator $\hat{f}(S)$ is applied to the sample $S$ (or
equivalently, depends on the set $S^*$) and we
 require that for all $\vecv$ and $u$, $\hat{f}(S(u,\vecv))\geq 0$
 (estimator is nonnegative), and for
  all $\vecv$, $\E_{u\sim U[0,1]}[ \hat{f}(S(u,\vecv))]= f(\vecv)$
  (estimator is unbiased).

The main motivation for MEP comes from coordinated
  sampling, which dates back to Brewer et al \cite{BrEaJo:1972}, and
  was extensively applied in Statistics and Computer Science~\cite{Saavedra:1995,ECohen6f,Ohlsson:2000,Rosen1997a,Broder:CPM00,BRODER:sequences97,CK:sigmetrics09,LiChurchHastie:NIPS2008,multiw:VLDB2009}.
With coordinated sampling, we can think of our data set as a nonnegative
  matrix $\{v_{ij}\}$.  Each row
  corresponds to a different time or location and each column to a
  particular key or feature.  The data set is sampled, for example,
  each row can be Reservoir sampled, or Probability Proportional to Size (PPS)
  sampled,  so only a small number of nonnegative entries are retained.
 With coordinated sampling, the samples of different rows utilize the
 same randomization (this can be achieved by applying a hash function
 to each key ID).  As a result,  we obtain the property that the
 samples of different rows are more similar when the data is.   This
 property is also known as the Locality
 Sensitive Hashing (LSH)  property.  One of the benefits of
 coordinating the samples  of rows  is
 that it facilitates tighter estimates of many important functions of multiple
 rows, such as their similarity.

   We are interested in estimating {\em sum aggregates} from the sampled
  data.   A sum aggregate is a sum over selected keys (columns) of some function
  $f$ of the values the key assumes in one or more instances.

 An example of an aggregate that is of particular importance in data
 analysis is $L^p_p$ of
  two rows, which is the sum over keys of the exponentiated range
  function, $\range_p$, which is the absolute difference
  between their values in the two instances, raised to the power of
  $p>0$.  The $L_p$ distance, an extensively used distance measure,  is the $p$th root of $L_p^p$.
We studied estimation of $\range_p$ (over coordinated and independent
samples) in \cite{sdiff:KDD2014}.  

  We can estimate such sums by considering each key (column) separately.
When the samples of different rows are coordinated, we obtain a
simpler estimation problem for each key:  If $\vecv = (v_{1h},v_{2h},\ldots,v_{rh})$ are the values the
key $h$ assumes in different rows, we would like to estimate $f(\vecv)$ from the sample.
For example, to estimate $L_p^p$,  we  estimate
$f(\vecv)=\range_p(\vecv)$ for each selected key, and then sum the
estimates.  Note that in this framework, the estimate is $0$ on keys
for which we have no information, we therefore, similar to estimation from
a single set (row), we only need to actively
compute the estimator for keys that are sampled in at least one
row.  This allows the computation of the estimate of the sum aggregate
to be scalable.

   When the sample of different rows are {\em coordinated}, the
problem for each key is a MEP.  We are often interested in nonnegative $f$, which means that we favor
   nonnegative estimators $\hat{f}$.  Since our main application is
   estimating sums, and the variance will be high for a typical key
   (since we typically have no or little information on the values),
   unbiased estimators  are desirable.

Classic point estimation theory studies estimating the
parameter(s) $\theta$ of a distribution the data was drawn from, and 
estimators and quality measures of estimators are extensively studied since the
time of Gauss \cite{Lehmann:book}.    A {\em risk} function which
assigns cost for the deviation of the estimator from the true value is
used.  A popular risk function is the expected
squared error (which for unbiased estimators is the variance).
The risk, however, depends on the parameter.
Ideally, we would want a Uniform Minimum Variance Unbiased Estimator
(UMVUE), which minimizes variance for all parameter values.  In
reality, for most estimation problems, including monotone estimation in general, a UMVUE
does not exist. 
Instead, we desire to
somehow balance the risk across possible parameter values.  In
estimation theory, the main approaches either assume a distribution over $\theta$ and
minimizing the average risk, or a Minimax estimator which minimizes
the maximum risk.  Either way,  we are interested in an admissible (Pareto optimal) estimator,
which means that it can not be strictly improved, attaining strictly lower risk for
some parameters without increasing the risk on some others.

In our MEP formulation, which is suited for data analysis from samples, there are no
distribution assumptions.  Instead,  we estimate an arbitrary
function of the data $\vecv$.   As we explained earlier, we only consider unbiased
nonnegative estimators, since we are interested in sum estimators and
nonnegative functions. The risk function we work with is the
squared error, which since we only consider unbiased estimators, is
the same as the variance.

 For MEP, a UMVUE generally does not exist and we similarly aim to
 ``balance'' performance over the domain.
  We recently proposed a relative  ``risk'' measure, inspired by {\em competitive
  analysis} in theoretical computer science \cite{BorodinLinialSaks:JACM1992}.  For each vector $\vecv$, we consider
 the minimum variance attainable by a nonnegative unbiased estimator.  We
then define the variance ratio of an estimator to be the maximum over 
$\vecv\in {\bf V}$, of the expectation of the square of the estimator
to the minimum possible by an unbiased nonnegative estimator.  That
is, instead of considering the absolute deviation, we compare
performance,point wise,  to the best possible. 
 The advantage of competitiveness over the more classic measures is
 that it does not require distribution assumptions on the data and
 also
that it captures performance in a way that is loosens up the
dependence on the magnitude of $f$.  A property  that is not critical in the parameter
estimation setting but is important in our setting.

  Surprisingly perhaps, we learned that we can characterize this
  point-wise variance optimum
  for any MEP and data $\vecv\in {\bf V}$ \cite{CKsharedseed:2012,sorder:PODC2014}.  Moreover,
for any MEP for which an unbiased nonnegative estimator with finite variances exists,
there exists an estimator with a constant  ratio   \cite{CKsharedseed:2012}.
In \cite{sorder:PODC2014} we presented a particularly natural estimator,
the \L\ estimator, that is guaranteed to be $4$ competitive.  We also
showed that the ratio of $4$ is tight for the \L\ estimator:  For any $\epsilon>0$, 
there is an MEP for which the \L\ estimator has ratio $\geq 4-\epsilon$.

  Our previous work \cite{CKsharedseed:2012,sorder:PODC2014}, left open two
  natural questions on variance competitive estimator constructions
  for MEPs.   The first is mostly of theoretical interest, the second is of both
  theoretical and   practical interest.

\begin{quote}
$\bullet$  What is the smallest possible competitive ratio which can be 
guaranteed for all queries and data domains (for which an unbiased 
nonnegative estimator with bounded variances exist) ?
We refer to this ratio as the {\em universal ratio} for monotone estimation.

$\bullet$   For a specific MEP, can we construct an estimator with minimum ratio for this MEP ?
We refer to such an estimator as {\em optimally competitive}.
\end{quote}

 We partially address the first question in Section~\ref{alphaL:sec} by presenting a parameterized construction
 of estimators, which are valid for any MEP for which an unbiased
 nonnegative estimator with finite variances exist. 
This family of estimators, which we name the $\alpha \L$ estimators,
has a parameter $\alpha\geq 1$.  When
$\alpha=1$, we obtain the \L\ estimator of \cite{sorder:PODC2014},
which is 4-competitive.  The \L\ estimators follows the lower bound of
the optimal range, having a ratio closer to $1$ for data $\vecv$ where
$f(\vecv)$ is small.  For $\alpha>1$, the estimator lies in the middle
of the optimal range of estimators.
We show that for $\alpha=1.5$ we
obtain an upper 
bound  of $27/8 \approx 3.38$ on the ratio for any MEP.
There fore we obtain a tighter upper bound which strictly improves
over the previous bound of $4$ obtained by  our \L\ estimator.

In Section \ref{lower:sec},
we obtain a lower bound on the universal ratio by first devising a
method to construct an optimally competitive estimator for MEP over a
finite domain.  We then 
conduct a computer 
search over certain function families on finite domains.  For these instances 
we computed the instance-optimal competitive ratio. 
The highest ratio we encountered in our search was 1.44, which gives a lower 
bound of 1.44 on the universal ratio.


\section{Preliminaries} \label{prelim:sec}

 We review some material that is necessary for our presentation.
For a set $Z\subset {\bf V}$,  we
define $\underline{f}(Z) = \inf\{ f(v) \mid v\in Z\}$ to be
the infimum of $f$ on $Z$.
For an outcome $S(u,\vecv)$, we
use the notation $\underline{f}(S) \equiv
\underline{f}^{(\vecv)}(u) \equiv \underline{f}(S^*)$ for the infimum
of $f$ on all data vectors $S^*$consistent with the outcome.

From monotonicity of the sampling scheme, it follows that 
$\forall \vecv,\ \underline{f}^{(\vecv)}(u)$ is 
monotone non increasing in $u$.
It is also not hard to see that any unbiased and nonnegative estimator
$\hat{f}$ must satisfy
\begin{align}  
\forall \vecv, \forall \rho, \int_\rho^1 \hat{f}(u,\vecv)du \leq
\underline{f}^{(\vecv)}(\rho) \ .\label{nonneg:eq} 
\end{align}

The lower bound function $\underline{f}^{(\vecv)}$,  and its lower hull
$H^{(\vecv)}_f$, can be used to determine the existence
of estimators with certain properties~\cite{CKsharedseed:2012}:
\begin{eqnarray}
\bullet && \text{$\exists$  unbiased nonnegative $f$ estimator}
 \iff \label{nec_reqa}\\
&& \forall \vecv\in {\bf V},\,
\lim_{u \rightarrow 0^+} \underline{f}^{(\vecv)}(u) = f(\vecv)\
. \label{nec_req}\\
\bullet && \text{If $f$ satisfies \eqref{nec_req},}\,  
 \text{$\exists$ unbiased nonnegative estimator with finite variance for $\vecv$} \nonumber\\
&&  \iff \int_0^1 \bigg(\frac{d H^{(\vecv)}_f(u)}{du} \bigg)^2 du  < \infty \ .  \label{bounded_var_shared_nec_req}
\end{eqnarray}

\medskip
\noindent
We work with {\em partial specifications} $\hat{f}$ of (nonnegative and unbiased)
estimators.  The specification is for a set of outcomes that is closed
to increased $u$:  For all $\vecv$, there is $\rho_v$, so that
$S(u,\vecv)$ is specified if and only if $u> \rho_v$.
If $\rho_v=0$,  the estimator is
{\em fully specified} for $\vecv$.

The partial specification is nonnegative, and we also require that for
any $\vecv$, the estimate values on the specified part never exceed
$f(\vecv)$, which from \eqref{nonneg:eq},  is clearly a necessary condition for extending the
specification to a nonnegative unbiased estimator.
We established in \cite{CKsharedseed:2012} that
if a MEP satisfies \eqref{nec_req} (has a nonnegative unbiased
estimator), then any  partially specified estimator
can be extended to an unbiased nonnegative estimator.  

 Our derivations
of estimators in \cite{CKsharedseed:2012,sorder:PODC2014} utilize
partial specifications: We express the estimate value on an outcome as a
function of the estimate values of all ``less-informative'' outcomes
(those with larger $u$).  
The specification is such that the estimate
on an outcome is selected to be ``optimal'' in some respect.  
In particular, we can precisely consider an optimal choice of
$\hat{f}(S)$ with respect to a particular consistent vector
$\vecv\in S^*$.

Given a partially specified estimator $\hat{f}$ so that $\rho_v>0$ and
$M=\int_{\rho_v}^1 \hat{f}(u,\vecv)du$, 
a {\em $\vecv$-optimal extension} is
an extension which is fully specified for $\vecv$ and, among all such
extensions, minimizes
variance for $\vecv$.
  The $\vecv$-optimal  extension is
defined on outcomes $S(u,\vecv)$ for $u\in (0,\rho_v]$ and minimizes
$\int_0^{\rho_v} \hat{f}(u,\vecv)^2 du$ subject to 
$\int_0^{\rho_v} \hat{f}(u,\vecv) du= f(\vecv)-M$ (unbiasedness), 
$\forall u,  \hat{f}(u,\vecv) \geq 0$ (nonnegativity), and
$\forall u,  \int_u^{\rho_v} \hat{f}(x,\vecv)dx \leq
\underline{f}^{(\vecv)}(u)-M$ (necessary nonnegativity for other
data).
At the point $\rho_v$, the {\em $\vecv$-optimal estimate} is
\begin{equation}
\lambda(\rho,\vecv,M) =\inf_{0\leq \eta < \rho}
\frac{\underline{f}(\eta,\vecv)-M}{\rho-\eta}\ .\label{lambdaMdef}
\end{equation}
 For the outcome $S(\rho,v)$, we can also consider the range of
 optimal estimates (with respect to $M$).  The infimum and supremum of this range are
\begin{eqnarray}
\lambda_U(S,M)
&=& \sup_{\vecz\in S^*(\rho,\vecv)}
\lambda(\rho,\vecz,M)\ \label{lambdaULdef}\\
\lambda_L(S,M)
&=& \inf_{\vecz\in S^*(\rho,\vecv)}
\lambda(\rho,\vecz,M)  = \frac{\underline{f}(\rho,\vecv)-M}{\rho}\label{lambdaMLdef}
\end{eqnarray}
Estimators that are outside the range with finite probability can not
be (unbiased and nonnegative) admissible, that is, they can be
strictly improved.

 For $\rho_v\in (0,1]$ and $M\in [0,\underline{f}^{(\vecv)}(\rho_v)]$, we
define the function
$\hat{f}^{(\vecv,\rho_v,M)}:(0,\rho_v]\rightarrow R_+$ as the solution of
\begin{equation}
\hat{f}^{(\vecv,\rho_v,M)}(u) =\inf_{0\leq \eta < u}
\frac{\underline{f}^{(\vecv)}(\eta)-M-\int_u^{\rho_v}\hat{f}^{(\vecv,\rho_v,M)}(u)du}{\rho-\eta}\ . 
\end{equation}
Geometrically, the function $\hat{f}^{(\vecv,\rho_v,M)}$  is the
negated derivative of the lower hull of the lower bound function
$\underline{f}^{(\vecv)}$ on $(0,\rho_v)$ and the point $(\rho_v,M)$.  
\begin{theorem}  \label{voptlh} \cite{CKsharedseed:2012}
Given a partially specified
estimator $\hat{f}$ so that $\rho_v>0$ and
$M=\int_{\rho_v}^1 \hat{f}(u,\vecv)du$, then
$\hat{f}^{(\vecv,\rho_v,M)}$ is the unique (up to equivalence)
$\vecv$-optimal extension of $\hat{f}$.
\end{theorem}

The {\em $\vecv$-optimal} estimates are the minimum variance extension of
the empty specification. We use $\rho_v=1$ and $M=0$ and obtain
 $\hat{f}^{(\vecv)}\equiv \hat{f}^{(\vecv,1,0)}$.
$\hat{f}^{(\vecv)}$ is the solution of
\begin{equation}  \label{PWopt}
\hat{f}^{(\vecv)}(u) =\inf_{0\leq \eta < u}
\frac{\underline{f}^{(\vecv)}(\eta)-\int_u^1 \hat{f}^{(\vecv)}(u)du}{\rho-\eta}\ ,
\end{equation}
which is the negated slope of the lower hull of the lower bound function
$\underline{f}^{(\vecv)}$.

\smallskip
\noindent
{\bf Variance competitiveness}~\cite{CKsharedseed:2012} of an
estimator is defined with respect to the expectation of the square.
An estimator $\hat{f}$ is {\em $c$-competitive} if
$$\forall \vecv,\, \int_0^1 \bigg(\hat{f}(u,\vecv)\bigg)^2du \leq c \inf_{\hat{f}'}\int_0^1 \bigg(\hat{f}'(u,\vecv)\bigg)^2du ,$$
where the infimum is over all unbiased nonnegative estimators of $f$.
An estimator that minimizes the expectation of the square also
minimizes the expected squared error.  When unbiased, it minimizes the variance.

\section{Upper bound on the universal ratio} \label{alphaL:sec}



 We define the family of 
$\alpha$\L\ estimators, with respect to a parameter $\alpha\geq 1$. 
This family extends the definition of the \L\ estimator we presented
in \cite{sorder:PODC2014}, which is the special case of  $\alpha=1$.  
The \L\ estimator is defined by searching for an estimate that is the
minimum possible in the ``optimal range'' of admissible estimators.
As a result, the estimator is ``optimized'' that is, has variance that
is close to the minimum possible for data vectors with a smaller
$f(\vecv)$.  The \L\ estimator is also the unique {\em monotone}
estimator, meaning that for any $\vecv$ the estimate is never lower on
a more informative outcomes.

For larger $\alpha$, the $\alpha$\L\ estimator gives more weight to the less
informative outcomes. 
More precisely,  the $\alpha \L$ estimator, $\hat{f}^{(\alpha L)}(x,\vecv)$, for random seed value $x$ and on outcomes 
  consistent with some fixed data  $\vecv$, is the solution of the 
  integral   equation, $\forall \vecv$, $\forall x\in (0,1]$,
\begin{equation} \label{alphaeq}
\hat{f}^{(\alpha L)}(x,\vecv) = \frac{\alpha}{x}\bigg(\underline{f}^{(\vecv)}(x) - \int_x^1 
\hat{f}^{(\alpha L)}(u,\vecv) du\bigg)\ . 
\end{equation}
 
We assume here that the lower bound function satisfies 
 $\underline{f}^{(\vecv)}(1)=0$:  Otherwise, if we are interested in estimating 
 functions $f$  where this is not the case, we can 
shift the lower bound function by subtracting $\underline{f}^{(\vecv)}(1)$,
compute the estimator with respect to the shifted function, and then 
add back the constant $\underline{f}^{(\vecv)}(1)$ to the estimate. 
The expectation-of-square ratio computed for the shifted function can only be 
lower than the ratio obtained when $\underline{f}^{(\vecv)}(1)=0$. 
From \eqref{alphaeq}, we get 
$\hat{f}^{(\alpha L)}(1,\vecv)=\alpha  \underline{f}^{(\vecv)}(1)=0$. 

Similarly to the special case of the \L\ estimator we treated in \cite{sorder:PODC2014}, the $\alpha \L$ estimate value depends 
only on information available from the outcome, which is the values of 
the lower bound function and the estimate value on less informative 
outcomes.  Therefore, the estimates are consistently defined  across
the data domain.
We note that for $\alpha<1$,
  these estimators lie outside the optimal range on 
{\em every} outcome.  Therefore, the $\alpha \L$ estimator in this
case is dominated by the \L\ estimator and thus is not interesting.

For $\alpha>1$, the $\alpha \L$ estimators, which 
solve $\hat{f}(\rho,S) = \alpha \lambda_L(S)$  are not necessarily
in-range. 

To force the estimator to be in-range (which results in strict improvement) we can 
instead define it the solution of 
$\hat{f}(S) = \min\{\lambda_U,\alpha\lambda_L\}$.  
Unbiasedness and nonnegativity of $\alpha \L$ follow immediately then from being 
in-range \cite{sorder:PODC2014}, but also hold without the truncation to 
$\lambda_U$. 
The upper bound establish next on the competitiveness of the
$\alpha$\L\ estimators also applies to the definition without this truncation.

We establish the following:
\begin{theorem} \label{alphaL:thm}
The $\alpha\L$ estimator is $\frac{4\alpha^3}{(2\alpha-1)^2}$-competitive.  
The supremum of the ratio over instances is at least 
$\frac{4\alpha^2}{(2\alpha-1)^2}$. 
\end{theorem}


 Fixing the data $\vecv$, the lower bound function 
 $\underline{f}^{(\vecv)}(x)$ is bounded (upper bounded by 
  $f(\vecv)$ and lower bounded by $0$) and monotone 
  non-increasing and hence differentiable almost 
  everywhere.   We multiply \eqref{alphaeq} by $x$ and take a 
  derivative with respect to $x$ and obtain the first-order differential equation 
\begin{equation}  \label{dalphaL:eq}
x\frac{\partial \hat{f}(x,\vecv)}{\partial x}  - (\alpha-1) 
\hat{f}(x,\vecv) = \alpha \frac{\partial 
  \underline{f}(x,\vecv)}{\partial x}\ . 
\end{equation}
The solution is uniquely determined when we incorporate the 
initial condition $\hat{f}(1,\vecv)=0$:
\begin{equation} \label{soldalphaL:eq}
\hat{f}^{(\alpha L)}(x,\vecv)= - \alpha x^{\alpha-1}\int_x^1 
y^{-\alpha}\frac{\partial \underline{f}(y,\vecv)}{\partial y} dy \ . 
\end{equation}

 To study competitiveness, we can consider the estimate values and 
 the lower bound function with respect to a fixed data $\vecv$.   We therefore 
omit the reference to  $\vecv$  in the notation. 
For convenience, we define $g(x)= - \frac{\partial 
  \underline{f}(x,\vecv)}{\partial x} \geq 0$ and 
obtain the equation for $\hat{f}$ (with initial condition) and solution $\hat{f}_{\alpha,g}$:
\begin{eqnarray}
x\hat{f}'(x) - (\alpha-1) \hat{f}(x) &=& -\alpha g(x) \ ,
\hat{f}(1)=0 \label{gdalphaL:eq}\\
\hat{f}_{\alpha,g}(x) &=& \alpha x^{\alpha-1}\int_x^1 
y^{-\alpha} g(y)dy \label{soldalphaLg:eq}
\end{eqnarray}
  We now bound the ratio of 
  $\int_0^1 \hat{f}_{\alpha,g}(x)^2 dx$ to $\int_0^1 g(x)^2dx$.   
This corresponds to the ratio of the  
expectation of the square of the 
  $\alpha  \L$ estimator  to  $\int_0^1 g(x)^2dx=\int_0^1 \bigg(\frac{\partial 
    \underline{f}^{(\vecv)}(x)}{\partial x}\bigg)^2dx$.   
 When the lower bound 
  function is convex ($g(x)$ is monotone non-increasing),
  from Theorem~\ref{voptlh}, 
 $g(x)$
  are the $\vecv$-optimal estimates, and $\int_0^1 g(x)^2dx$ is the 
  minimum expectation 
  of the square for $\vecv$, over all unbiased nonnegative estimators. 


\begin{theorem}  \label{alphaLconvex:thm}
Let $g(x) \geq 0$ on $(0,1]$ be such that 
$\int_0^1 g(x)^2dx < \infty$.  For $\alpha \geq 1$, let $\hat{f}(x) 
\equiv \hat{f}_{\alpha,g}$ be the solution of  \eqref{gdalphaL:eq}. 
Then 
\begin{equation}
\int_0^1 \hat{f}(x)^2 dx \leq \bigg(\frac{2\alpha}{2\alpha-1}\bigg)^2 \int_0^1 
g(x)^2 dx\ . 
\end{equation}
\end{theorem}
\begin{proof}
Rearranging  \eqref{gdalphaL:eq},
 we obtain 
\begin{equation}  \label{subst:eq}
x\hat{f}'(x) = (\alpha-1) \hat{f}(x)  -\alpha g(x)\ . 
\end{equation}

{\scriptsize 
\begin{eqnarray}
(\hat{f}(x)^2 )' &=& 2\hat{f}(x)\hat{f}'(x) \, \implies \\
\hat{f}(x)^2 &=& -2 \int_x^1 \hat{f}(y)\hat{f}'(y) dy \, \implies \label{intbyparts:eq}\\
\int_0^1 \hat{f}(x)^2 dx &=& - 2 \int_0^1 \int_x^1 \hat{f}(y)\hat{f}'(y) 
dy dx = - 2 \int_0^1 
x \hat{f}'(x) \hat{f}(x) dx \label{doubleintorder:eq}\\
&=&  -2 \int_0^1 \hat{f}(x) \bigg( (\alpha-1) \hat{f}(x)  -\alpha g(x) \bigg) 
dx \label{usubst:eq}\\
&=& -2 (\alpha-1) \int_0^1 \hat{f}(x)^2 dx  + 2\alpha \int_0^1 
\hat{f}(x)g(x) dx \ . \label{frel:eq}  
\end{eqnarray}
}
We applied integration by parts to obtain \eqref{intbyparts:eq}, and 
then changed order of double  integration \eqref{doubleintorder:eq},
using the initial condition $\hat{f}(1)=0$, and reduced to a single integral. 
To obtain \eqref{usubst:eq}, we substituted \eqref{subst:eq}. 
Rearranging \eqref{frel:eq},  and using Cauchy-Schwarz inequality, we 
obtain 
{\scriptsize 
\begin{eqnarray}
 \int_0^1 \hat{f}(x)^2 dx &=& \frac{2\alpha}{2\alpha-1}
\int_0^1 \hat{f}(x) g(x) dx  \label{f2intfg:eq}\\
&\leq& \frac{2\alpha}{2\alpha-1} \sqrt{\int_0^1 \hat{f}(x)^2 
  dx}\sqrt{\int_0^1 g(x)^2 dx}\ . 
\end{eqnarray}
}
Finally, the claim of the theorem follows by dividing both sides by $\sqrt{\int_0^1 \hat{f}(x)^2 
  dx}$ and squaring. 
\end{proof}

 We now show that the expectation of the square of the $\alpha \L$ estimates 
with respect to a lower bound  function $\underline{f}(x)$ 
with lower hull $H(x)$, is bounded by $\alpha$ times the 
expectation of the square of the estimator computed with respect 
to the convex lower bound function $H(x)$.  The statement of 
 the theorem is in terms of the negated derivatives, $h(x)$ and $g(x)$,
 of $H(x)$ and $\underline{f}(x)$:

\begin{lemma}  \label{alphaLconvex:lemma}
Let $h(x) \geq 0$ be monotone non-increasing on $(0,1]$ such that 
$\int_0^1 h(x)^2dx < \infty$.  Define $H(x)\equiv \int_x^1 h(u)du$. 
Let $g(x)$ be such that the lower hull of $G(x)\equiv \int_x^1 g(u)du$
is equal to $H(x)$.  Then 
for $\alpha\in (1,2]$,
$$ \int_0^1 \hat{f}_{\alpha,g}(x)^2 dx \leq \alpha  \int_0^1 \hat{f}_{\alpha,h}(x)^2dx \ .$$
\end{lemma}
\begin{proof}
Let $\hat{f} \equiv \hat{f}_{\alpha,g}$ be the solution 
\eqref{soldalphaL:eq}  of  \eqref{gdalphaL:eq}.  From the proof 
of Theorem \ref{alphaLconvex:thm},  $\hat{f}$ satisfies 
\eqref{f2intfg:eq}. 
Substituting \eqref{soldalphaL:eq}  in \eqref{f2intfg:eq}, we obtain 
\begin{eqnarray}
\int_0^1 \hat{f}(x)^2 dx &=& \frac{2\alpha}{2\alpha-1}
\int_0^1 \hat{f}(x) g(x) dx \nonumber \\
&=& \frac{2\alpha^2}{2\alpha-1} \int_0^1 g(x) x^{\alpha-1}\int_x^1 
y^{-\alpha}g(y) dy dx\ . \label{lhsgf}
\end{eqnarray}
We have $\int_0^1 g(x) dx = \int_0^1 h(x) dx$ and for all $x\in (0,1]$,
$\int_0^x g(u) du \leq  \int_0^x h(u) du$. 

Consider the {\em defining points} of the hull $H$.  These are the points so that for all $g$ defining the same hull, we must have 
$\int_x^1 g(x)dx = \int_x^1 h(x)dx$.  It suffices to show that 
$\int_a^b \hat{f}_{\alpha,g}(x)^2 dx \leq  \alpha \int_a^b \hat{f}_{\alpha,h}(x)^2 dx$ between 
any two such points.  Moreover, it suffices to consider only intervals between such points (the discontinuities).  For such an interval $[a,b]$, the function $h$ must be fixed (a linear part of the hull). 
We have 
{\scriptsize 
\begin{eqnarray}
\lefteqn{\int_a^b \hat{f}_{\alpha,g}(x)^2 dx =}\\
&=&  \frac{2\alpha^2}{2\alpha-1}
\int_a^b g(x) x^{\alpha-1}\int_x^1 y^{-\alpha}g(y) dy  dx\\
&=& \frac{2\alpha^2}{2\alpha-1} \int_a^b g(x) 
x^{\alpha-1}\bigg(\int_x^b y^{-\alpha}g(y) dy+\int_b^1 y^{-\alpha}g(y) 
dy\bigg) dx\ . 
\end{eqnarray}
}

 Between any two defining points, $\int_a^b g(x)dx = \int_a^b h(x)dx$ and also 
$\int_a^x g(u) du \leq  \int_a^x h(u) du$. 
We now fix  $g(x)$ in the interval $[b,1]$ and the integral $B_g= \int_b^1 y^{-\alpha}g(y) dy$. Since both $b$ and $1$ are defining points of the hull,  the properties above, and monotonicity of $y^{-\alpha}$, imply that $B_g \leq B_h$. 

 It suffices to show that 
\begin{equation}\label{alpharatio}
\frac{\int_a^b \hat{f}_{\alpha,g}(x)^2 
   dx}{\int_a^b \hat{f}_{\alpha,h}(x)^2 dx} \leq \alpha\ . 
\end{equation}
The function $h(x)$ is constant on $(a,b)$. 
Let $h(x)=A$ on $(a,b)$.  To bound the ratio \eqref{alpharatio}, we 
separately consider and bound 
the ratio of $g$ to $h$ for each of two summands:  $\int_a^b g(x) x^{\alpha-1} B_g dx$ and 
$\int_a^b g(x) x^{\alpha-1} \int_x^b y^{-\alpha}g(y) dy dx$. 

For the first summand, we have 
\begin{align*}\int_a^b g(x) x^{\alpha-1} B_g dx 
\leq  B_g b^{\alpha-1}(b-a)A \leq B_h A (b^{\alpha} -a b^{\alpha-1})\\
\leq  B_h A (b^{\alpha} -a^{\alpha}) \ .\end{align*}
We have $\int_a^b h(x) x^{\alpha-1} B_h dx = B_h A (b^\alpha-a^\alpha)/\alpha$. 
We get that the ratio is at most $\alpha$.  

 We now consider the ratio of the second summand 
$$\int_a^b g(x) x^{\alpha-1} \int_x^b y^{-\alpha}g(y) dy dx\ ,$$  for $g$ and $h$. 

At the denominator, we have the expression for $h(x)$, which is 
\begin{eqnarray}
\lefteqn{\int_a^b h(x) x^{\alpha-1} \int_x^b y^{-\alpha}h(y) dy  dx=} \nonumber\\
&=&  A^2 \int_a^b x^{\alpha-1} \int_x^b y^{-\alpha} dy dx \nonumber \\
&=&  \frac{A^2}{\alpha-1} \int_a^b x^{\alpha-1} (x^{1-\alpha}
-b^{1-\alpha}) dx\\ &=&  \frac{A^2}{\alpha-1} \bigg( (b-a) 
-\frac{b^{1-\alpha}}{\alpha}(b^\alpha-a^\alpha) \bigg) \nonumber \\
 &=&  \frac{A^2}{\alpha-1} \bigg((b-a) 
 -\frac{b}{\alpha}(1-\left(\frac{a}{b}\right)^\alpha) \bigg) 
 \label{constA} \ . 
\end{eqnarray}

We now consider 
$\int_a^b g(x) x^{\alpha-1} \int_x^b y^{-\alpha}g(y) dy$


We approximate $g$ by a piecewise constant function, on $n$ pieces,
each containing $1/n$ of the mass.  The breakpoints are $a\equiv t_0 < t_1 
\cdots < t_n \equiv b$ satisfy $\int_a^{t_i} g(x) dx = i(b-a)A/n$. 
The breakpoints must satisfy  $t_i \geq a+ i(b-a)/n$. 
The fixed value in $(t_i,t_{i+1})$ is $W_i=\frac{(b-a)A}{n(t_{i+1}-t_i)}$
We have 
for $j>i$,
\begin{align*}
T_{ij} &\equiv \int_{t_i}^{t_{i+1}} g(x) x^{\alpha-1}  \int_{t_j}^{t_{j+1}} g(y) 
  y^{-\alpha} dy dx \\
&= \int_{t_i}^{t_{i+1}} W_i x^{\alpha-1}  \int_{t_j}^{t_{j+1}} W_j 
  y^{-\alpha} dy dx \\
&= W_i W_j  \frac{t_i^{\alpha-1} - t_{i+1}^{\alpha-1}}{\alpha}
  \frac{t_{j}^{-\alpha+1} -t_{j+1}^{-\alpha+1}}{\alpha-1}  \\
&= \frac{(b-a)^2 A^2}{n^2 \alpha (\alpha-1)}  \frac{(t_i^{\alpha-1} - t_{i+1}^{\alpha-1})(t_{j}^{-\alpha+1} -t_{j+1}^{-\alpha+1})}{(t_{i+1}-t_i)(t_{j+1}-t_j)}
\end{align*}

For $i$,
\begin{align*}
T_{ii} &\equiv \int_{t_i}^{t_{i+1}} g(x) x^{\alpha-1}  \int_{x}^{t_{i+1}} g(y) 
  y^{-\alpha} dy dx \\ &= W_i^2 \int_{t_i}^{t_{i+1}} x^{\alpha-1}
  \frac{x^{-\alpha+1} -t_{i+1}^{-\alpha+1}}{\alpha-1} \\ &= 
W_i^2\frac{\bigg( (t_{i+1}-t_i) -
  \frac{t_{i+1}^{-\alpha+1}}{\alpha}(t_{i+1}^\alpha-t_i^{\alpha})\bigg)}{\alpha-1}
\\
&= W_i^2\frac{\bigg( (t_{i+1}-t_i) -
  \frac{t_{i+1}}{\alpha}(1-
  \frac{t_{i}^\alpha}{t_{i+1}^{\alpha}})\bigg)}{\alpha-1} \\
&= \frac{(b-a)^2 A^2}{n^2 (\alpha-1)} \frac{\bigg( (t_{i+1}-t_i) -
  \frac{t_{i+1}}{\alpha}(1-
  \frac{t_{i}^\alpha}{t_{i+1}^{\alpha}})\bigg)}{(t_{i+1}-t_i)^2} 
\end{align*}

 The expression is $\sum_{i=0}^{n-1} \sum_{j=i}^{n-1} T_{ij}$.  We 
 need to show that for all $n$, the maximum over sequences $t$ is 
 bounded by $\alpha$ times \eqref{constA}.

  For $\alpha=2$, we obtain $T_{ii}=\frac{(b-a)^2 A^2}{2n^2}
  \frac{1}{t_{i+1}}$ and $T_{ij}= \frac{(b-a)^2 A^2}{2n^2}
  \frac{1}{t_j t_{j+1}}$.  The sum is maximized when all $t_i$ are at 
  their minimum value of $t_i a+ i(b-a)A/n$, which means all the $W_i$
  are equal to $A$. 

 More generally, for $\alpha\in (1,2]$, the partial derivatives of 
 $T_{ij}$ with respect to $t_i$, $t_j$, $t_{i+1}$, $t_{j+1}$,  and of 
 $T_{ii}$ with respect to $t_i$ and $t_{i+1}$,  are all negative. 
This means that the sum is maximized when $t_i$ are as small as 
possible, and we can use the same argument.

\ignore{We now observe that for all $g$, $\int_x^1 y^{-\alpha}g(y) dy \leq \int_x^1 
y^{-\alpha}h(y) dy$.  This is 
because $g(x)\equiv h(x)$ maximizes $\int_0^x g(x)dx$, for 
all $x$, amongst all qualifying functions $g$, and $y^{-\alpha}$ is decreasing. 
We now consider the function $m(x)=x^{\alpha-1}\int_x^1 y^{-\alpha}h(y) dy$
and argue that  for $\alpha\in [1,2]$,  since $h(x)$ is monotone 
non-increasing, the function  $m(x)$ is 
 monotone  non-increasing in $x$.  
Lastly, we consider $\int_0^1 g(x) m(x) dx$, since $m(x)$ is monotone 
non-increasing, reusing the argument above, the integral is maximized 
for $g(x)=h(x)$.  
We therefore obtain that \eqref{lhsgf}, and hence,
$\int_0^1 \hat{f}_{\alpha,g}(x)^2 dx$,  is maximized for $g=h$.}
\end{proof}

  Combining the results from Theorem~\ref{alphaLconvex:thm} and 
  Lemma~\ref{alphaLconvex:lemma}, we obtain that the $\alpha$L estimator is 
  $4\alpha^3/(2\alpha-1)^2$ competitive.  This expression is minimized 
for $\alpha=1.5$, where we get a competitive ratio $27/8=3.375$. 

 To conclude the proof of Theorem \ref{alphaL:thm}
we need to show that for any $\epsilon>0$ there are instances where $\alpha \L$ has ratio at least $\frac{4\alpha^2}{(2\alpha-1)^2}-\epsilon$:
\begin{lemma}
The supremum of the ratio of the $\alpha \L$ estimator is $\geq \frac{4\alpha^2}{(2\alpha-1)^2}$. 
\end{lemma}
\begin{proof}
 Consider the function $f(v)=1-v^p$ ($p\in (0.5,1]$), where $v\in [0,1]$. 
For data $v=0$, the lower bound function is $1-v^p$ and is 
square integrable for $p\in (0,5,1]$. 
Since the lower bound function is convex, 
the $0$-optimal estimates are $\hat{f}^{(0)}(x)= \underline{f}(x)'=
p/x^{1-p}$. 
The optimal expectation of the square  is $\frac{p^2}{2p-1}$. 

  The $\alpha \L$ estimator is 
$\hat{f}^{(\alpha L)}(x) = \frac{\alpha 
  p}{\alpha-p}(x^{p-1}-x^{\alpha-1})$. 
The expectation of the square is 

$$\int_0^1 \hat{f}^{(\alpha L)}(x)^2 dx = \frac{\alpha^2 
  p^2}{(\alpha-p)^2}\bigg(
\frac{1}{2\alpha-1}+\frac{1}{2p-1}-\frac{2}{\alpha+p-1}\bigg)\ .$$
Simplifying, we obtain the ratio of $\int_0^1 \hat{f}^{(\alpha L)}(x)^2 dx$ to the optimum:
$$\frac{2\alpha^2}{(2\alpha-1)(\alpha+p-1)}\ .$$

 Fixing $\alpha$, we look at the supremum over $p\in (0.5,1]$ of this 
 ratio,
which is obtained for $p\rightarrow 
0.5^+$ and is equal to $\frac{4\alpha^2}{(2\alpha-1)^2}$. 
\end{proof}
We obtain ratio $\geq 4$ for $\alpha=1$ (the \L\ estimator) 
 and $\geq 16/9$ for $\alpha=2$.

 \section{Lower bound on the universal ratio} \label{lower:sec}
We start with a simple example of a MEP where any (nonnegative
unbiased) estimator has ratio that is at least $10/9$.   This gives a
lower bound of $10/9$ on the universal ratio.

 The data domain has 3  points: ${\bf V}=\{0,0.5,1\}$ and the function is 
$f(0)=2$, $f(0.5)=1$, and $f(1)=0$.   
The sampling scheme is such that data $v\in {\bf V}$, is sampled
$\iff$  $u< v$.  That is, if $u<v$ then $S^*=\{v\}$ and otherwise,
$S^*=[0,u)\cap {\bf V}$.
The lower bound function for $v=1$ is 
 $\underline{f}^{(1)}(u) \equiv 0$, for $v=0.5$ is $\underline{f}^{(0.5)}(u)=1$ for $u\in (0,1)$ and for  $v=0$, we have  is $\underline{f}^{(0)}(u)=2$ for  $u\in (0,0.5]$ and $\underline{f}^{(0)}(u)=1$ for $u\in (0.5,1)$.  
The $v$-optimal estimates for each of 
$v\in \{0,0.5,1\}$ are fixed $\hat{f}^{(v)}(u)\equiv f(v)$ for 
$u\in (0,1)$.  
The optimal expectation of the square is therefore $f(v)^2$. 

Any variance optimal 
nonnegative unbiased  estimator must be $0$ when the data is $1$. 
When the 
data is $\{0,0.5\}$, the estimator must 
have the same fixed value $y \in [0,2]$ for $x\in (0.5,1)$  and a different fixed 
value (determined by $y$, $v$,  and unbiasedness) when 
$v\in\{0.5,1\}$.  This value is equal to 
$2-y$ when $v=0.5$ and to  $4-y$ when $v=1$. 
(since information is the same on all these outcomes, variance is 
minimized when the estimate is the same).  
The respective expectation of the square, as a function of $y$, is accordingly $y^2/2+ (2-y)^2/2 =
y^2-2y+2$ for $v=0.5$ and is $y^2/2 + (4-y)^2/2 = y^2+8-4y$ for $v=0$. 
The two ratios are respectively $y^2-2y+2$ for $v=0.5$ and $y^2/4+2-y$
for $v=1$.  The competitive ratio is minimized by $y$ which minimizes 
the maximum of $y^2-2y+2$ and $y^2/4+2-y$.  The maximum is minimizes when 
 $y=4/3$.  The corresponding ratio of this estimator is $10/9$.

\subsection{Computer search for a tighter lower bound}
Using a computer program we computed the optimal ratio on MEPs
on discrete domains which included thousands of points. 
We obtained instances where any estimator must have ratio that is 
at least  $1.44$.  

Providing more detail, we considered discrete one-dimensional 
domain ${\bf V}=\{i/n\}$ for $i=0,\ldots,n$.  The sampling scheme we use is
PPS sampling of $v$:  For 
  $u\sim U[0,1]$,  we ``sample'' $v$  if and only if $v\geq u$.  The
  respective monotone sampling scheme has
$S^*(u,v)=\{v\}$ when $u\leq v$ and $S^*(u,v)=[0,u)\cap {\bf V}$ otherwise.

We are then interested in estimating 
$f(v)=1-v^p$ for $p\in (0,1]$ and estimating 
$f(v)=(1-v)^p$ for $p>1$.  It is easy to verify that on this finite
domain,
unbiased nonnegative estimators with finite variances exist for all 
$p$ and $n$.     

Any (nonnegative unbiased) admissible estimator must  have a very particular structure.
The value $\hat{f}(v,u)$ is the same for all $v<u$.  For $u<v$, when
we know $v$ exactly,  the
estimate is determined by the values for $\hat{f}(v,u)$ for
$u>v$ and unbiasedness.   Moreover, an admissible estimator is also fixed in
each interval $X_i= (i/n,(i+1)/n]$, since the information we have, in
terms of $S^*$,  within the interval is the same.

  It therefore suffices to consider the values of the estimator on the
  $n$ points $i/n$, and ensure for unbiasedness that for any $v$, the
  integral over $u>v$ does not exceed $f(v)$.

  We first implemented a subroutine which (attempts to) construct a $c$-competitive estimator for
  a particular $c$.  
We also compute the $v$-optimal estimate for any
$v\in {\bf V}$.
The subroutine considers the intervals $X_i$ in decreasing $i$ order.  At each step,
we use the maximum estimate so that the ratio on affected data points
(those consistent with $v\leq i/n$ remains below $c$.  
We can then compute the full estimate for $u\leq i/n$ for the point
$v=i/n$ and test its competitiveness.
If there is no $c$-competitive estimator, that is,
the input choice of $c$ was too low, our subroutine  reveals that and
stops.  Otherwise, it finds a $c$ competitive estimator.

 We apply this subroutine in a binary 
search, looking for the
minimum value $c$ for which the
subroutine succeeds in building an estimator.  This allows us to
approximate or tightly lower bound, the optimal ratio for this MEP.
 The highest ratio we found on the MEPs we examined 
was 1.44. This implies a lower bound of 1.44 on the universal ratio.

  Finally, we note that this construction of the optimally-competitive
  estimator only applies with certain simple family of functions.  It
  would be interesting to come up with a general construction.
The particular function $1-v^p$ is interesting, since the \L\
estimator has a ratio 
which approaches its worst-case ratio of 4 \cite{sorder:PODC2014}.
The particular function $(1-v)^p$ is also interesting.  It is a
special case of the exponentiated range,
$\range_p(\vecv)=|v_1-v_2|^p$  which is the basis of
Manhattan and Euclidean distance estimation \cite{sdiff:KDD2014} with
PPS sampling.
In fact, our construction yields an optimally competitive ratio
of $\approx 1.204$ for $p=1$ and of $\approx 1.35$ for $p=2$, whereas the ratio of the \L\
estimator is respectively $2$ ($p=1$) and $2.5$ ($p=2$).


{\small
\bibliographystyle{plain}
\bibliography{cycle} 

\begin{thebibliography}{10}

\bibitem{BorodinLinialSaks:JACM1992}
A.~Borodin, N.~Linial, and M.~E. Saks.
\newblock An optimal on-line algorithm for metrical task system.
\newblock {\em J. ACM}, 39(4):745--763, 1992.

\bibitem{BrEaJo:1972}
K.~R.~W. Brewer, L.~J. Early, and S.~F. Joyce.
\newblock Selecting several samples from a single population.
\newblock {\em Australian Journal of Statistics}, 14(3):231--239, 1972.

\bibitem{BRODER:sequences97}
A.~Z. Broder.
\newblock On the resemblance and containment of documents.
\newblock In {\em Proceedings of the Compression and Complexity of Sequences},
  pages 21--29. IEEE, 1997.

\bibitem{Broder:CPM00}
A.~Z. Broder.
\newblock Identifying and filtering near-duplicate documents.
\newblock In {\em Proc.of the 11th Annual Symposium on Combinatorial Pattern
  Matching}, volume 1848 of {\em LNCS}, pages 1--10. Springer, 2000.

\bibitem{ECohen6f}
E.~Cohen.
\newblock Size-estimation framework with applications to transitive closure and
  reachability.
\newblock {\em J. Comput. System Sci.}, 55:441--453, 1997.

\bibitem{sdiff:KDD2014}
E.~Cohen.
\newblock Distance queries from sampled data: Accurate and efficient.
\newblock In {\em KDD}. ACM, 2014.
\newblock full version: {\tt http://arxiv.org/abs/1203.4903}.

\bibitem{sorder:PODC2014}
E.~Cohen.
\newblock Estimation for monotone sampling: Competitiveness and customization.
\newblock In {\em PODC}. ACM, 2014.
\newblock full version {\tt http://arxiv.org/abs/1212.0243}.

\bibitem{CK:sigmetrics09}
E.~Cohen and H.~Kaplan.
\newblock Leveraging discarded samples for tighter estimation of multiple-set
  aggregates.
\newblock In {\em ACM SIGMETRICS}, 2009.

\bibitem{CKsharedseed:2012}
E.~Cohen and H.~Kaplan.
\newblock What you can do with coordinated samples.
\newblock In {\em The 17th. International Workshop on Randomization and
  Computation (RANDOM)}, 2013.
\newblock full version: {\tt http://arxiv.org/abs/1206.5637}.

\bibitem{multiw:VLDB2009}
E.~Cohen, H.~Kaplan, and S.~Sen.
\newblock Coordinated weighted sampling for estimating aggregates over multiple
  weight assignments.
\newblock {\em Proceedings of the VLDB Endowment}, 2(1--2), 2009.
\newblock full version: {\tt http://arxiv.org/abs/0906.4560}.

\bibitem{Lehmann:book}
E.~L. Lehmann.
\newblock {\em Theory of point estimation}.
\newblock Wadsworth, California, 1983.

\bibitem{LiChurchHastie:NIPS2008}
P.~Li, , K.~W. Church, and T.~Hastie.
\newblock One sketch for all: Theory and application of conditional random
  sampling.
\newblock In {\em NIPS}, 2008.

\bibitem{Ohlsson:2000}
E.~Ohlsson.
\newblock Coordination of pps samples over time.
\newblock In {\em The 2nd International Conference on Establishment Surveys},
  pages 255--264. American Statistical Association, 2000.

\bibitem{Rosen1997a}
B.~Ros{\'e}n.
\newblock Asymptotic theory for order sampling.
\newblock {\em J. Statistical Planning and Inference}, 62(2):135--158, 1997.

\bibitem{Saavedra:1995}
P.~J. Saavedra.
\newblock Fixed sample size pps approximations with a permanent random number.
\newblock In {\em Proc. of the Section on Survey Research Methods, Alexandria
  VA}, pages 697--700. American Statistical Association, 1995.

\end{thebibliography}
}


\end{document}